\documentclass[12pt,notitlepage,a4paper]{article}
\usepackage{graphicx}
\usepackage{amsmath}
\usepackage{epsfig}

\begin{document}

\title{\textbf{Counterfactual errors and state reduction in relativistic quantum
physics}}
\author{Jon Eakins and George Jaroszkiewicz \\
School of Mathematical Sciences, University of Nottingham, \\
University Park, Nottingham NG7 2RD, UK}
\date{\today }
\maketitle

\begin{abstract}
We use the laws of relativistic physics to show that classically motivated
counterfactual statements are inadequate when discussing the principles of
quantum physics and that EPR style arguments against state reduction are
incorrect.
\end{abstract}

In classical logic, counterfactual statements are statements of the form:
\emph{``if $P$ were true (which is in fact known not to be the case) then $Q$
would be true''}. Such statements are often acceptable in classical
mechanics because one of its cardinal unstated principles is that a
classical, objective reality exists regardless of whether or not anything
has been observed.

When it comes to quantum mechanics, however, the greatest care must be taken
to adopt the right way of thinking. Physicists generally have sound
intuition when it comes to classical mechanics, but it is all too easy to
use this intuition inappropriately when discussing quantum mechanics. One
such argument concerns state reduction and is the subject of this note.
State reduction is somewhat unfashionable currently, particularly from the
point of view of decoherence theory, which asserts that only continuous in
time Schr\"{o}dinger unitary evolution occurs, contrary to the state
reduction postulate of von Neumann \cite{VON-NEUMANN:55}. Clearly, it is of
importance to our understanding of the universe to have a sound view of this
issue, because it has implications at every level of physics, ranging from
sub-atomic physics to cosmology. If we accept incomplete or misleading
arguments against state reduction, there remains the possibility that we
have misunderstood the nature of quantum reality.

Perhaps the most famous example of inadmissable classical thinking applied
to quantum mechanics is to be found in the Einstein-Podolsky-Rosen (EPR)
paper \cite{EPR}, which attempted to give an argument for the incompleteness
of quantum mechanics. Although the EPR argument was refuted immediately by
Bohr \cite{BOHR-35}, the sort of thinking used by EPR has persisted, with
the consequence that basic errors continue to be made in the interpretation
of quantum mechanics. One of the difficulties in accepting Bohr's
counterattack is that it requires us to think of entangled quantum systems
holistically. This is classically counter-intuitive because it involves
non-locality, which is anathema in classical mechanics. Since there still
appears to be confusion in the recent literature surrounding the
interpretation of the EPR ``paradox'', the aim of this note is to point out
the dangers of applying classical counterfactual reasoning to quantum
mechanics. Specifically, we shall refute the Stanford (Encyclopedia of
Philosophy) argument against state reduction \cite{KRIPS-99} by using its
own line of reasoning and the principles of relativistic physics. There is
no need to appeal to recent developments in quantum information theory.

As with the Stanford argument, we discuss a particular ``thought
experiment''. In our case, however, we are confident that it could be
carried out in actuality, because we avoid making any counterfactual
statements, which, by definition, can never be proved. The experiment is
equivalent to the one discussed in the EPR paper \cite{EPR} and in \cite
{KRIPS-99}, confronting nonlocality in relativity (i.e., separation in
space) with the von Neumann state reduction postulate \cite{VON-NEUMANN:55}
in quantum mechanics (i.e., wave function collapse). It involves a spin-zero
state $\Psi $ of an electron-positron system created by some apparatus $O$
(see Figure 1). Considering only the spins of the particles, $\Psi $ is an
entangled element of a tensor product Hilbert space $\mathcal{H}=\mathcal{H}%
_{e}\otimes \mathcal{H}_{p}$, where $\mathcal{H}_{e}$ and $\mathcal{H}_{p}$
represent the spin degrees of freedom of the electron and positron
respectively. A spin zero state is of the form
\begin{equation}
\Psi =e_{\mathbf{n}}^{+}p_{\mathbf{n}}^{-}-e_{\mathbf{n}}^{-}p_{\mathbf{n}%
}^{+},
\end{equation}
where $\mathbf{n}$ is any unit three-vector in physical space. Here $e_{%
\mathbf{n}}^{+}$ is an eigenstate of electron spin measured along the
direction $\mathbf{n}$ with eigenvalue $+1$, $p_{\mathbf{n}}^{-}$ is an
eigenstate of positron spin along $\mathbf{n}$ with eigenvalue $-1$, and so
on. The state $\Psi $ is invariant to spatial rotations and therefore we are
at liberty to choose any direction in physical space for the unit vector $%
\mathbf{n}$.

Suppose now that $A$ and $B$ are two spatially separated observers at rest
in some inertial frame $F$, such that $A$ tests for electron spin whilst $B$
tests for positron spin. These tests are carried out over sufficiently small
intervals of space and time so that we can identify them with events (points
in spacetime), also labelled by $A$ and $B$ respectively. In frame $F$,
mutually orthogonal spatial axes have been previously chosen and denoted by
the unit vectors $\mathbf{i}, \mathbf{j}$ and $\mathbf{k}$ respectively, and
there is no ambiguity about this concerning $A$ or $B$. The test $\Sigma _{%
\mathbf{k}}^{e}$ employed by $A$ tests for electron spin along the $\mathbf{k%
}$-direction and therefore has two possible outcomes, $e_{\mathbf{k}}^{+}$
and $e_{\mathbf{k}}^{-},$ whilst the test $\Sigma _{\mathbf{j}}^{p}$
employed by $B$ tests for positron spin along the $\mathbf{j}$ direction and
also has two possible outcomes, $p_{\mathbf{j}}^{+}$ and p$_{\mathbf{j}%
}^{-}$.
\begin{center}
\begin{figure}[t]
\centerline{\epsfxsize=5.0in \epsfbox{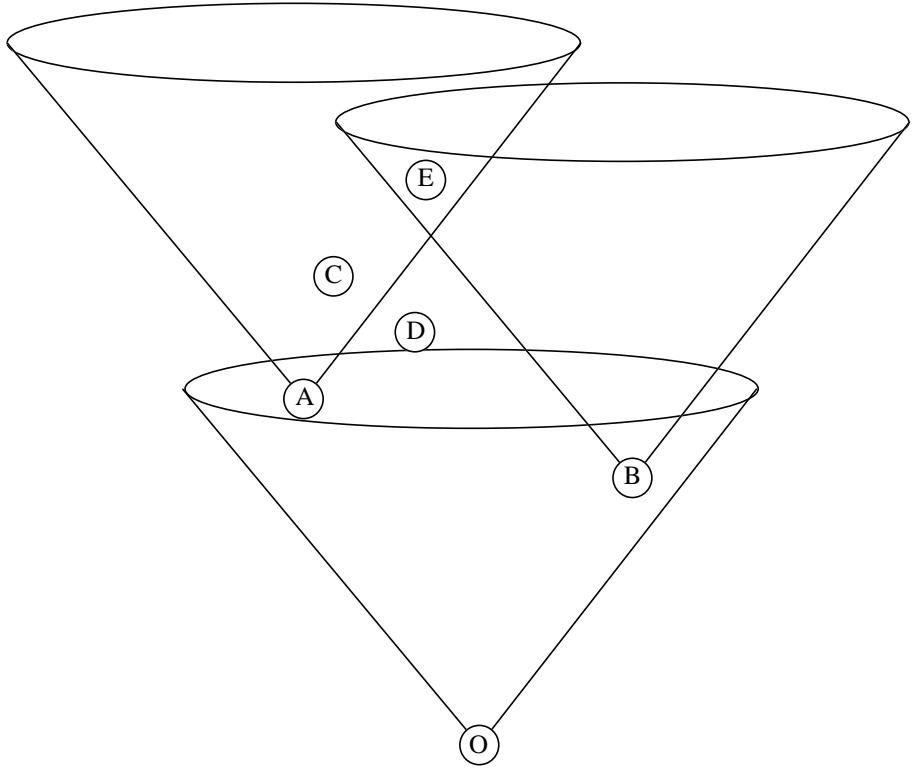}}
\caption{$A$ tests for electron spin whereas $B$ tests for positron spin. $C$
and $D$ are excluded by charge conservation from testing for positron spin
given $B$, whereas $E$ is a possible second test for positron spin.}
\end{figure}
\end{center}
The argument for the inconsistency of the state reduction postulate given in
\cite{KRIPS-99} appears quite convincing and goes as follows. The tests
carried out by $A$ and $B$ on their respective particles are conducted with
a spacelike interval between them, so they cannot interfere with each other.
Suppose at time $t_{A}$, $A$ finds the electron to be in state $e_{\mathbf{k}%
}^{+}$. Then by conservation of total angular momentum, $A$ would deduce
that, for any subsequent time $t > t_{A}$, the positron state has collapsed
into the anti-correlated state $p_{\mathbf{k}}^{-}$. Essentially, $A$'s
measurement of the electron's spin is equivalent to an indirect ideal
measurement \cite{PERES:93} of the positron's spin.

But suppose now that $B$ had actually tested for positron spin at time $t_{B}
$ just before time $t_{A}$, according to observers in frame $F$, i.e., just
before $A$ had performed their test $\Sigma _{\mathbf{k}}^{e}$ on the
electron. $B$'s test throws the positron into an eigenstate of $\Sigma _{%
\mathbf{j}}^{p}$, i.e., into either $p_{\mathbf{j}}^{+}$ or else $p_{\mathbf{%
j}}^{-}$, according to the reduction postulate. Because of the spacelike
interval between them, Einstein locality (the principle of local causes \cite
{PERES:93}) implies that the two tests $\Sigma _{\mathbf{k}}^{e},\Sigma _{%
\mathbf{j}}^{p}$ could not possibly interfere with each other. But this
means that, at any time $t > t_{A}$, the positron is in state $p_{\mathbf{k}%
}^{-}$ because of $A$'s test, whereas it is in either of the states $p_{%
\mathbf{j}}^{+}$, $p_{\mathbf{j}}^{-}$ because of $B$'s test. Since the
positron cannot be in two definite states at once, the Stanford conclusion
is that state reduction is inconsistent.

This argument is equivalent to the original EPR argument and is incorrect
for the same reasons \cite{BOHR-35}. The error lies with the unfettered
application of counterfactual statements to quantum mechanical situations.
In the EPR case, direct tests of position and momentum cannot be carried out
on the same particle simultaneously. In the case of the Stanford experiment,
$A$'s assertion that the positron is in the state $p_{\mathbf{k}}^{-}$ cannot
be verified by $B$, because $B$ has already tested the positron (with outcome
$p_{\mathbf{j}}^{+} $ or $p_{\mathbf{j}}^{-})$. \emph{In quantum physics, it
very much matters what is actually done, rather than speculated on}. As it
stands, $A$'s belief about the positron's state after $A$'s observation of the
electron is a counterfactual one (with the twist that it relates to a
non-existent future experiment by $B$, rather than to a non-existent past one)
and therefore it does not have the logical status of $B$'s valid knowledge
about the positron's state after time zero.

In quantum mechanics, it is meaningless to talk about a state without any
reference to any test of that state. $A$'s statement about the positron spin
would have to be tested by a \emph{second} test for positron spin, over and
above the test at $B$, in order for it (i.e., $A$'s statement) to have physical
significance. However, any additional test for positron spin at events such as
$C$ or $D$ which are outside the forwards lightcone of $B$ could not be
completed, because this would violate charge conservation, which is believed to
hold absolutely (there would be at least one inertial frame in which \emph{two}
positrons existed at different places simultaneously in such a case). The only
possible region of spacetime where a second test of the positron's spin could
be carried out consistent with known physics is at some event $E$ which is in
the region of overlap of the forwards lightcones from both $A$ and $B$. But
then it could not be argued that $E$ was testing only the state of the positron
as prepared indirectly by $A$, because $B$ could causally influence $E$.
Therefore, $A$'s view of the positron state is inadmissible and the Stanford
argument against wave function collapse itself collapses.

It is easy to see that the probability of $E$ finding the positron in state $%
p_{\mathbf{k}}^{+}$, given that $A$ had found the electron in state $e_{%
\mathbf{k}}^{+}$ and that $B$ had tested for positron spin along $\mathbf{j}$%
, is one half and not the value zero as predicted by $A$
on the basis of angular momentum conservation. The conclusions
we draw from this is that the greatest care must be taken in the interpretation
of quantum mechanics, particularly where state reduction is concerned. We
conjecture that the state reduction concept will never lead to a paradox in
quantum physics, provided proper care is taken to eliminate counterfactual
errors.

Finally, we remark that the state reduction concept need not be invoked if
only one test (or measurement) of a quantum system is being discussed. In
such a case Schr{\"{o}}dinger (unitary) evolution can be used, up to the
point of measurement, plus the Born probability interpretation, without
reference to state reduction. That is why decoherence calculations are valid
under such circumstances. However, if a second test is to be performed
subsequently, as in the case of event E in our thought experiment above,
then state reduction prior to that second test must be invoked.

\end{document}